\documentclass[preprint]{ptephy_v1}

\preprintnumber{KYUSHU-HET-290} 

\usepackage{graphics}

\usepackage{amsmath} 
\usepackage{amsthm} 
\usepackage{url} 
\usepackage{stmaryrd}
\usepackage{tikz}
\usepackage{physics}

\numberwithin{equation}{section}

\allowdisplaybreaks

\begin{document}

\title{Action of the axial $U(1)$ non-invertible symmetry on the 't~Hooft
line operator: A simple argument}


\author{Yamato Honda}

\author{Soma Onoda}

\author{Hiroshi Suzuki}

\affil{Department of Physics, Kyushu University, 744 Motooka, Nishi-ku,
Fukuoka 819-0395, Japan}








\begin{abstract}%
Employing the modified Villain lattice formulation of the axion quantum
electrodynamics, we present an alternative and much simpler derivation of the
conclusion of~Ref.~\cite{Honda:2024sdz} that the sweep of the axial $U(1)$
non-invertible symmetry operator over the (non-genuine) gauge invariant
't~Hooft line operator with an integer magnetic charge does not leave any
effect. The point is that such a 't~Hooft line can be represented by a boundary
of a (non-topological) defect that is invariant under the axial transformation
on the axion field.
\end{abstract}

\subjectindex{B01,B04,B31}

\maketitle

\section{Introduction and summary}
\label{sec:1}
The study in~Refs.~\cite{Choi:2022jqy,Cordova:2022ieu} pointed out the
possibility that the axial $U(1)$ rotation of the fermion field with fractional
angles, although it suffers from the ordinary axial $U(1)$ anomaly, may be
understood as a generalized form of the symmetry~\cite{Gaiotto:2014kfa} (see
Refs.~\cite{Schafer-Nameki:2023jdn,Bhardwaj:2023kri,Shao:2023gho} for reviews),
the non-invertible symmetry~\cite{Aasen:2016dop,Bhardwaj:2017xup,Chang:2018iay,Thorngren:2019iar,Komargodski:2020mxz,Koide:2021zxj,Choi:2021kmx,Kaidi:2021xfk,Hayashi:2022fkw,Choi:2022zal,Kaidi:2022uux,Roumpedakis:2022aik,Bhardwaj:2022yxj,Cordova:2022ieu,Choi:2022jqy,Bhardwaj:2022lsg,Karasik:2022kkq,GarciaEtxebarria:2022jky,Choi:2022fgx,Yokokura:2022alv,Nagoya:2023zky,Anber:2023mlc}. In~Ref.~\cite{Honda:2024sdz},
the present authors studied how the symmetry operator of the axial $U(1)$
non-invertible symmetry acts on the (non-genuine) 't~Hooft line operator by
modeling the axial anomaly by the axion and employing the modified Villain
lattice formulation of $U(1)$ gauge theory~\cite{Sulejmanpasic:2019ytl} (see
also~Ref.~\cite{Gorantla:2021svj}), which allows a straightforward introduction
of the dual $U(1)$ gauge field. In~Ref.~\cite{Honda:2024sdz}, it was concluded
that the sweep of the symmetry operator over the 't~Hooft line operator does
not leave any effect. This conclusion appears inequivalent with the phenomenon
concluded in~Refs.~\cite{Choi:2022jqy,Cordova:2022ieu} that the sweep leaves
non-trivial traces. Although we believe that under assumptions made
in~Ref.~\cite{Honda:2024sdz}, our conclusion is correct, the argument
in~Ref.~\cite{Honda:2024sdz} might be somewhat complicated and, in particular,
it might appear that the conclusion depends crucially on a miraculous
cancellation between the effect of the dressing factor in the 't~Hooft line and
the would-be Witten effect~\cite{Witten:1979ey} arising from the axion
topological coupling.

In this paper, we present an alternative and much simpler derivation of the
conclusion in~Ref.~\cite{Honda:2024sdz}, although the basic assumptions are
common to both. Considering the basic importance of this issue in possible
applications of the axial $U(1)$ non-invertible symmetry, especially in the
form of the selection rules, we believe that it is useful to do this. In this
paper, we show that the conclusion in~Ref.~\cite{Honda:2024sdz} can be derived
from the following observation: The (non-genuine) 't~Hooft line operator
considered in~Ref.~\cite{Honda:2024sdz} can be realized by a background
2-cocycle gauge field, which couples to the (non-conserved) electric 2-cocycle
current~$\star f$. Since this coupling is not influenced by the axial
transformation on the axion field, $\phi\to\phi+\alpha$, the symmetry operator
does not cause any effect on the 't~Hooft line. This simple observation
explains the above miraculous cancellation. Also, we hope that the present
simple argument gives an idea that our conclusion, although it is shown by
employing a particular lattice action, is quite insensitive on the details of
the lattice action one adopts. The extension of the present argument to the
case of the action of the 1-form non-invertible symmetry in the axion quantum
electrodynamics (QED)~\cite{Choi:2022fgx,Yokokura:2022alv}, the action of the
1-form non-invertible symmetry operator on the axion string, would be an
interesting problem.\footnote{Since the (non-genuine) gauge invariant axion
string can be represented by a background 1-cochain gauge field introduced in
the kinetic term of the axion as we will see in~Sect.~\ref{sec:2.5}, and
the 1-form non-invertible symmetry is associated with the shift of the
gauge potential, repeating a similar argument as that in the present paper, we
expect that the sweep of the symmetry operator over the axion string does not
leave any effect.}

\section{The action of the axial $U(1)$ non-invertible symmetry operator}
\label{sec:2}
\subsection{Lattice action}
We take the lattice action of~Ref.~\cite{Honda:2024sdz} (Eq.~(2.24) there) as
an explicit starting point for our argument. It is based on the modified
Villain formulation of the $U(1)$ lattice gauge
theory~\cite{Sulejmanpasic:2019ytl}. After the integration by parts on the
periodic hypercubic lattice~$\Gamma$, the lattice action can be written
as:\footnote{We denote the hypercube, the cube, the plaquette, and the link
on~$\Gamma$ by $h$, $c$, $p$, and~$l$, respectively. The $p$-cochain, the cup
product~$\cup$, and the coboundary operator~$\delta$ on~$\Gamma$ are, roughly
speaking, lattice analogues of the $p$-form, the wedge product~$\wedge$, and
the exterior derivative~$\mathrm{d}$ on~$T^4$, respectively. For precise
definitions of the $p$-cochain, the cup product, the higher cup
product~$\cup_1$, the coboundary operator and the ``Hodge dual'' $\star$ on the
hypercubic lattice, see Refs.~\cite{Chen:2021ppt,Jacobson:2023cmr}. On the cup
product, the coboundary operator~$\delta$ satisfies the Leibniz rule,
$\delta(\alpha\cup\beta)=\delta\alpha\cup\beta+(-1)^p\alpha\cup\delta\beta$,
where $\alpha$ is a $p$-cochain. The cup product is not commutative on the
cochain level but the commutator is expressed by the higher cup as
$(-1)^{pq}\beta\cup\alpha=\alpha\cup\beta+(-1)^{p+q}[\delta(\alpha\cup_1\beta)
-\delta\alpha\cup_1\beta-(-1)^p\alpha\cup_1\delta\beta]$, where $\beta$ is a
$q$-cochain.}
\begin{align}
   S&=\sum_{h\in\Gamma}\biggl\{
   \frac{1}{2g_0^2}f\cup\star f+i\Tilde{a}\cup\delta z
   +\frac{\mu^2}{2}\partial\phi\cup\star\partial\phi
   +i\delta\ell\cup\chi
\notag\\
   &\qquad\qquad{}
   -i\frac{\mathrm{e}^2}{8\pi^2}\left[
   \partial\phi\cup(\mathrm{CS})-4\pi^2\phi\cup P_2(z)\right]\biggr\},
\label{eq:(2.1)}
\end{align}
where we assume that the electric $U(1)$ charge~$\mathrm{e}$ is an even integer
as~Ref.~\cite{Honda:2024sdz}. In this expression,
\begin{equation}
   f:=\delta a+2\pi z,
\label{eq:(2.2)}
\end{equation}
is the $U(1)$ field strength 2-cochain ($a\in\mathbb{R}$ is the $U(1)$ gauge
potential 1-cochain, $\delta$ is the coboundary operator and $z\in\mathbb{Z}$
is an integer 2-cochain field). $\Tilde{a}\in\mathbb{R}$ is a 1-cochain
Lagrange multiplier field and the integration over~$\Tilde{a}$ imposes the
Bianchi identity~$\delta z=0$; $\Tilde{a}$ can be regarded as the dual $U(1)$
gauge potential. For the axion scalar field~$\phi$, $\partial\phi$ is the
combination defined by
\begin{equation}
   \partial\phi:=\delta\phi+2\pi\ell,
\label{eq:(2.3)}
\end{equation}
where $\phi\in\mathbb{R}$ is a 0-cochain and $\ell\in\mathbb{Z}$ is an integer
1-cochain field. $\chi\in\mathbb{R}$ in~Eq.~\eqref{eq:(2.1)} is another
2-cochain auxiliary field which imposes the another Bianchi
identity~$\delta\ell=0$. The lattice $U(1)$ Chern--Simons form $(\mathrm{CS})$
is defined by~\cite{Jacobson:2023cmr}:\footnote{The motivation for this
definition may be understood as follows: In the present lattice formulation,
$f\cup f$ is not a total divergence. One can make this total divergence up to
$8\pi^2\mathbb{Z}$ by adding terms containing~$\delta z$ as
\begin{equation}
   f\cup f-4\pi a\cup\delta z+2\pi f\cup_1\delta z
   =\delta(\mathrm{CS})+4\pi^2 P_2(z).
\label{eq:(2.4)}
\end{equation}
}
\begin{equation}
   (\mathrm{CS}):=
   a\cup f+2\pi z\cup a+2\pi a\cup_1\delta z.
\label{eq:(2.5)}
\end{equation}
Finally, the Pontryagin square is defined by
\begin{equation}
   P_2(z):=z\cup z+z\cup_1\delta z.
\label{eq:(2.6)}
\end{equation}

In the modified Villain formulation~\cite{Sulejmanpasic:2019ytl}, the lattice
action and observables must be invariant under the following three gauge
transformations: The $\mathbb{Z}^{(1)}$ gauge transformation,
\begin{equation}
   a\to a+2\pi m,\qquad z\to z-\delta m,
\label{eq:(2.7)}
\end{equation}
where $m\in\mathbb{Z}$ is a 1-cochain. The $\mathbb{Z}^{(0)}$ gauge
transformation,
\begin{equation}
   \phi\to\phi+2\pi k,\qquad\ell\to\ell-\delta k,
\label{eq:(2.8)}
\end{equation}
where $k\in\mathbb{Z}$ is a 0-cochain. The $\mathbb{R}^{(0)}$ gauge
transformation,
\begin{equation}
   a\to a+\delta\lambda,\qquad z\to z,
\label{eq:(2.9)}
\end{equation}
where $\lambda\in\mathbb{R}$ is a 0-cochain. The first two gauge invariances
remove redundancies in the above description of the lattice $U(1)$ gauge theory
with a $2\pi$ periodic scalar field. The last one is the ordinary local $U(1)$
gauge symmetry.

In the above form of the lattice action~\eqref{eq:(2.1)}, the invariance under
the $\mathbb{Z}^{(0)}$ gauge transformation~\eqref{eq:(2.8)} is manifest,
because $\partial\phi$ in~Eq.~\eqref{eq:(2.3)} is a manifestly
$\mathbb{Z}^{(0)}$ gauge invariant combination. It can be seen
that~\cite{Honda:2024sdz} the lattice action~\eqref{eq:(2.1)} is invariant also
under other two gauge transformations,\footnote{%
One finds that, under the $\mathbb{Z}^{(1)}$ gauge
transformation~\eqref{eq:(2.7)},
\begin{align}
   (\mathrm{CS})
   &\to(\mathrm{CS})
   -2\pi\delta(m\cup a)+8\pi^2m\cup z
   -4\pi^2\delta(m\cup_1z)+4\pi^2\delta m\cup_1z-4\pi^2\delta m\cup m,
\notag\\
   P_2(z)&\to P_2(z)
   -2\delta m\cup z-\delta(\delta m\cup_1z)+\delta m\cup\delta m,
\label{eq:(2.10)}
\end{align}
and therefore, up to surface terms,
\begin{align}
   &\left[\partial\phi\cup(\mathrm{CS})-4\pi^2\phi\cup P_2(z)\right]
\notag\\
   &\to
   \left[\partial\phi\cup(\mathrm{CS})-4\pi^2\phi\cup P_2(z)\right]
   +8\pi^2\phi\cup m\cup\delta z
   -4\pi^2\delta\ell\cup m\cup a+8\pi^3\mathbb{Z}.
\label{eq:(2.11)}
\end{align}
Under the $\mathbb{R}^{(0)}$ gauge transformation~\eqref{eq:(2.9)},
up to surface terms,
\begin{align}
   &\left[\partial\phi\cup(\mathrm{CS})-4\pi^2\phi\cup P_2(z)\right]
\notag\\
   &\to
   \left[\partial\phi\cup(\mathrm{CS})-4\pi^2\phi\cup P_2(z)\right]
\notag\\
   &\qquad{}
   +4\pi\left(\phi\cup\delta\lambda-2\pi\ell\cup\lambda\right)\cup\delta z
   +2\pi\delta\ell\cup
   \left[
   \lambda\cup\delta a
   +2\pi\left(\lambda\cup z+z\cup\lambda+\lambda\cup_1\delta z\right)
   \right].
\label{eq:(2.12)}
\end{align}
} provided that we set the transformation laws of the auxiliary fields as,
under the $\mathbb{Z}^{(1)}$ gauge transformation,
\begin{equation}
   \Tilde{a}\to\Tilde{a}+\mathrm{e}^2\phi\cup m,\qquad
   \chi\to\chi-\frac{\mathrm{e}^2}{2}m\cup a,
\label{eq:(2.13)}
\end{equation}
and, under the $\mathbb{R}^{(0)}$ gauge transformation,
\begin{equation}
   \Tilde{a}\to\Tilde{a}+\frac{\mathrm{e}^2}{2\pi}
   \left(\phi\cup\delta\lambda-2\pi\ell\cup\lambda\right),\qquad
   \chi\to\chi+\frac{\mathrm{e}^2}{4\pi}
   \left[\lambda\cup\delta a+2\pi\left(\lambda\cup z+z\cup\lambda
   +\lambda\cup_1\delta z\right)
   \right].
\label{eq:(2.14)}
\end{equation}

\subsection{'t Hooft line operator as a defect boundary}
We introduce the 't~Hooft line operator in the present system as follows. We
couple a background 2-cochain gauge field~$B\in\mathbb{Z}$ to the electric
2-cochain current, $\star f$. For this, we change the kinetic term of the
gauge field in~Eq.~\eqref{eq:(2.1)} as
$f\cup\star f\to(f-2\pi B)\cup\star(f-2\pi B)$ and correspondingly define
\begin{align}
   S^B&:=\sum_{h\in\Gamma}\biggl\{
   \frac{1}{2g_0^2}\left(f-2\pi B\right)\cup\star\left(f-2\pi B\right)
   +i\Tilde{a}\cup\delta z
   +\frac{\mu^2}{2}\partial\phi\cup\star\partial\phi
   +i\delta\ell\cup\chi
\notag\\
   &\qquad\qquad{}
   -i\frac{\mathrm{e}^2}{8\pi^2}\left[
   \partial\phi\cup(\mathrm{CS})-4\pi^2\phi\cup P_2(z)\right]
   \biggr\}.
\label{eq:(2.15)}
\end{align}
The Poincar\'e dual of~$B$ defines a 2-dimensional surface
(i.e.\ defect)~$\mathcal{R}$. If $\delta B\neq0$, the surface~$\mathcal{R}$ has
the boundary~$\gamma$, $\partial\mathcal{R}=\gamma$, along which
$\delta B\neq0$. Then the gauge coupling~\eqref{eq:(2.15)} defines a 't~Hooft
line operator along~$\gamma$. Having these facts in mind, we set
\begin{equation}
   B=-\mathrm{q}\delta_2[\mathcal{R}],
\label{eq:(2.16)}
\end{equation}
where $\mathrm{q}$ is an integer (magnetic charge) and $\delta_2[\mathcal{R}]$
is the 2-cochain delta function whose support is~$\mathcal{R}$. This implies
that
\begin{equation}
   \delta B=-\mathrm{q}\delta_3[\gamma],
\label{eq:(2.17)}
\end{equation}
where $\delta_3[\gamma]$ is the 3-cochain delta function along~$\gamma$.

In the present axion QED, $\star f$ is not conserved under the equation of
motion, $\delta\star f\neq0$, and thus the partition function is not invariant
under the 1-form gauge transformation on~$B$, $B\to B+\delta\omega$.
Correspondingly, the Poincar\`e dual of~$B$, the surface~$\mathcal{R}$ is not
topological, i.e.\ the partition function with the Boltzmann weight~$e^{-S^B}$
depends on a precise shape of~$\mathcal{R}$. Nevertheless, we may still
introduce such a defect and define a (non-genuine) 't~Hooft line operator.
Note that, since the combination~$(f-2\pi B)\cup\star(f-2\pi B)$ is manifestly
invariant under the gauge transformations
in~Eqs.~\eqref{eq:(2.7)}--\eqref{eq:(2.9)}, the above method defines an object
invariant under those gauge transformations.\footnote{The continuum counterpart
of~Eq.~\eqref{eq:(2.1)} possesses an electric $\mathbb{Z}_{\mathrm{e}^2}$ 1-form
(invertible) symmetry~\cite{Hidaka:2020iaz,Hidaka:2020izy} and one can consider
a 't~Hooft line operator with a fractional magnetic charge as the boundary of
the corresponding symmetry operator. Such a fractionally-charged magnetic
monopole worldline cannot be considered in the present lattice formulation with
the integer~$z$. One would have to consider the ``excision
method''~\cite{Abe:2023uan} to incorporate such a fractionally-charged 't~Hooft
line; the excision method has not completely been developed for 4D gauge
theory.}

If we change the variable~$z$ in~$S^B$~\eqref{eq:(2.15)} as~$z\to z+B$, we may
remove the background gauge field~$B$ in the gauge kinetic term. Instead, the
other parts in the action~\eqref{eq:(2.15)} produce new factors. Under the
shift~$z\to z+B$, we find that the Boltzmann weight changes as
\begin{equation}
   e^{-S^B}\to e^{-S}T_{\mathrm{q}}(\gamma),
\label{eq:(2.18)}
\end{equation}
where the 't~Hooft line operator~$T_{\mathrm{q}}(\gamma)$ is given by
\begin{align}
   &T_{\mathrm{q}}(\gamma)
\notag\\
   &=
   \exp\Biggl[
   -i\sum_{h\in\Gamma}\biggl(
   \left(\Tilde{a}-\frac{\mathrm{e}^2}{2\pi}\phi\cup a\right)\cup\delta B
   +\frac{\mathrm{e}^2}{2\pi}
   \left(\phi\cup f-2\pi\ell\cup a\right)\cup B
\notag\\
   &\qquad\qquad\qquad\qquad{}
   +\frac{\mathrm{e}^2}{4\pi}
   \bigl\{
   \phi\cup
   \left[
   \delta\left(f\cup_1 B\right)-2\pi\delta z\cup_1B
   +2\pi B\cup_1\delta z+2\pi P_2(B)
   \right]
\notag\\
   &\qquad\qquad\qquad\qquad\qquad\qquad{}
   +2\pi\ell\cup
   \left[\delta\left(a\cup_1 B\right)-\delta a\cup_1B\right]
   \bigr\}\biggr)\Biggr]
\notag\\
   &=\exp\left\{
   -i\sum_{h\in\Gamma}\left[
   \left(\Tilde{a}-\frac{\mathrm{e}^2}{2\pi}\phi\cup a\right)\cup\delta B
   +\frac{\mathrm{e}^2}{2\pi}
   \left(\phi\cup f-2\pi\ell\cup a\right)\cup B
   -\frac{\mathrm{e}^2}{4\pi}\partial\phi\cup\left(f\cup_1 B\right)\right]
   \right\}.
\label{eq:(2.19)}
\end{align}
In the last equality, we have used the constraints $\delta z=-\delta B$
and~$\delta\ell=0$\footnote{One can further introduce the axion string for
which~$\delta\ell\neq0$ along a 2-surface by the method in~Sect.~\ref{sec:2.5}.
Then the resulting hybrid system of the 't~Hooft line and the axion string is
defined gauge invariantly.} imposed under the integration over $\Tilde{a}$
and~$\chi$ and the relations, $B\cup B=\delta B\cup_1B=B\cup_1\delta B=0$,
which follow from the explicit form of~$B$ in~Eq.~\eqref{eq:(2.16)}. Although
we also have the factor,
$\exp[-i\pi\mathrm{e}^2\sum_{h\in\Gamma}\ell\cup(z\cup_1B)]$, this is unity when
the $U(1)$ charge~$\mathrm{e}$ is an even integer\footnote{We assume this
condition throughout this paper as per~Ref.~\cite{Honda:2024sdz}.} and can be
neglected. The expression~\eqref{eq:(2.19)} with~Eq.~\eqref{eq:(2.16)}, up to
the last gauge invariant
factor~$\exp[i\sum_{h\in\Gamma}(\mathrm{e}^2/4\pi)\partial\phi\cup(f\cup_1B)]$,%
\footnote{This gauge invariant factor does not affect the argument
in~Ref.~\cite{Honda:2024sdz}.} coincides with the 't~Hooft line operator
studied in~Ref.~\cite{Honda:2024sdz} (Eq.~(2.26) there). We thus observe that
the non-genuine gauge invariant 't~Hooft line operator
in~Ref.~\cite{Honda:2024sdz}, which possesses an integer magnetic
charge\footnote{We believe that this is also the case for the 't~Hooft line
operator in~Refs.~\cite{Choi:2022jqy,Cordova:2022ieu} as indicated in
Eqs.~(2.28) and~(2.29) of~Ref.~\cite{Choi:2022jqy} and Eq.~(28)
of~Ref.~\cite{Cordova:2022ieu}.} can be realized as the defect introduced
as~Eqs.~\eqref{eq:(2.15)} and~\eqref{eq:(2.16)}.\footnote{It can be seen that
our 't~Hooft line operator, although it possesses an associated surface
operator, is a charged object of the magnetic $U(1)$ symmetry in the present
system.}

\subsection{Axial $U(1)$ Ward--Takahashi identity}
Next, we define the expectation value in the sub-sector by:\footnote{%
We assume that, divided by the gauge volumes, the integration measure is
defined by
\begin{equation}
   \int\mathrm{D}[\Tilde{a}]\mathrm{D}[\phi]\mathrm{D}[\ell]
   \mathrm{D}[\chi]
   :=\prod_{l\in\Gamma}\left[\int_{-\pi}^\pi\mathrm{d}\Tilde{a}(l)\right]
   \prod_{s\in\Gamma}\left[\int_{-\pi}^\pi\mathrm{d}\phi(s)\right]
   \prod_{l\in\Gamma}\left[\sum_{\ell(l)=-\infty}^\infty\right]
   \prod_{p\in\Gamma}\left[\int_{-\pi}^\pi\mathrm{d}\chi(p)\right].
\label{eq:(2.20)}
\end{equation}}
\begin{equation}
   \left\langle\dotsb\right\rangle_{\mathrm{s}}^B
   :=\int\mathrm{D}[\Tilde{a}]\mathrm{D}[\phi]\mathrm{D}[\ell]
   \mathrm{D}[\chi]\,e^{-S^B}\dotsb.
\label{eq:(2.21)}
\end{equation}
Note that the gauge fields ($a$ and~$z$) are not yet integrated in this
expression. We then consider a constant shift of the integration
variable~$\phi$ in a 4-dimensional region~$\mathcal{V}_4\subset\Gamma$,
\begin{equation}
   \phi(s)\to
   \begin{cases}\phi(s)+\alpha&\text{for $s\in\mathcal{V}_4$},\\
   \phi(s)&\text{otherwise}.\\
   \end{cases}
\label{eq:(2.22)}
\end{equation}
We assume that $\mathcal{V}_4$ contains the 2-surface~$\mathcal{R}$
in~Eq.~\eqref{eq:(2.16)} as~Fig.~\ref{fig:1}.
\begin{figure}[htbp]
\centering
\begin{tikzpicture}[scale=0.5]
\draw[thick](7,3.9,0) circle(1.7 and 0.6);
  \draw[thick] (0,5,0) ..controls (8,3,0) and (4,7,0).. (10,5,0) ..controls (10,4,-3) and (10,6,-6).. (10,5,-8) .. controls (4,6,-8) and (8,4,-8) .. (0,5,-8) ..controls (0,6,-6) and (0,4,-3).. (0,5,0);
  \node at (8,5,-4) {$\mathcal{M}_3'$};
  \draw[thick] (0,0,0) ..controls (8,-2,0) and (4,2,0).. (10,0,0) ..controls (10,-1,-3) and (10,1,-6).. (10,0,-8) .. controls (4,1,-8) and (8,-1,-8) .. (0,0,-8) ..controls (0,1,-6) and (0,-1,-3).. (0,0,0);
  \node at (8,0,-4) {$\mathcal{M}_3$};
  \node at (4.1,3.6,0) {$\mathcal{V}_4$};
  \node at (9.2,3.8,0) {$\gamma$};
  \node at (6.98,3.93,0) {$\mathcal{R}$};
\end{tikzpicture}
\caption{Schematic view of our setup.}
\label{fig:1}
\end{figure}
This shift of the axion field corresponds to the axial $U(1)$ rotation of the
rotation angle~$\alpha$. We assume that the operator~$\dotsb$
in~Eq.~\eqref{eq:(2.21)} is invariant under this shift.
Under~Eq.~\eqref{eq:(2.22)}, the lattice action~\eqref{eq:(2.15)} changes by
\begin{align}
   S^B&\to
   S^B-\frac{i}{2}\alpha
   \left\{\sum_{c\in\partial\mathcal{V}_4}
   \left[\star j_5-\frac{\mathrm{e}^2}{4\pi^2}(\mathrm{CS})\right]
   -\mathrm{e}^2\sum_{h\in\mathcal{V}_4}P_2(z)\right\},
\label{eq:(2.23)}
\end{align}
where $j_5$ is the axial $U(1)$ 1-cochain current,
\begin{equation}
   j_5:=-2i\mu^2\partial\phi.
\label{eq:(2.24)}
\end{equation}
Since the integration measure~\eqref{eq:(2.20)} has an invariant meaning
under~Eq.~\eqref{eq:(2.22)},\footnote{That is, we may start with the
measure~$\int_{-\pi+\alpha}^{\pi+\alpha}\mathrm{d}\phi(s)$ for~$s\in\mathcal{V}_4$
instead of~Eq.~\eqref{eq:(2.20)} because of the $\mathbb{Z}^{(0)}$ gauge
invariance.} we have the identity,
\begin{equation}
   \left\langle\dotsb\right\rangle_{\mathrm{s}}^B
   =\left\langle
   \exp\left(
   \frac{i}{2}\alpha
   \left\{\sum_{c\in\partial\mathcal{V}_4}
   \left[\star j_5-\frac{\mathrm{e}^2}{4\pi^2}(\mathrm{CS})\right]
   -\mathrm{e}^2\sum_{h\in\mathcal{V}_4}P_2(z)\right\}\right)
   \dotsb\right\rangle_{\mathrm{s}}^B,\qquad
   \delta z=0.
\label{eq:(2.25)}
\end{equation}
Note that, as we have explicitly indicated here, the integration over the
auxiliary field~$\Tilde{a}$ in~Eq.~\eqref{eq:(2.21)} imposes the constraint
$\delta z=0$.\footnote{Under~$\delta z=0$, $P_2(z)$ is simply $z\cup z$.}

\subsection{Final steps}
We now assume that in~Eq.~\eqref{eq:(2.25)} the boundary of~$\mathcal{V}_4$
consists of two 3-surfaces,
\begin{equation}
   \partial\mathcal{V}_4=\mathcal{M}_3'\cup(-\mathcal{M}_3).
\label{eq:(2.26)}
\end{equation}
See~Fig.~\ref{fig:1}. We then substitute $\dotsb$ in~Eq.~\eqref{eq:(2.25)} by
\begin{equation}
   \exp\left\{
   \frac{i}{2}\alpha
   \sum_{c\in\mathcal{M}_3}
   \left[\star j_5-\frac{\mathrm{e}^2}{4\pi^2}(\mathrm{CS})\right]
   \right\}.
\label{eq:(2.27)}
\end{equation}
This is possible, because it can be seen that $\star j_5$ on~$\mathcal{M}_3$
is invariant under~Eq.~\eqref{eq:(2.22)}. Thus, we have
\begin{align}
   &\left\langle
   \exp\left\{
   \frac{i}{2}\alpha
   \sum_{c\in\mathcal{M}_3}
   \left[\star j_5-\frac{\mathrm{e}^2}{4\pi^2}(\mathrm{CS})\right]
   \right\}
   \right\rangle_{\mathrm{s}}^B.
\notag\\
   &=
   \exp\left[
   -\frac{i}{2}\alpha\mathrm{e}^2
   \sum_{h\in\mathcal{V}_4}P_2(z)\right]
   \left\langle
   \exp
   \left\{\frac{i}{2}\alpha
   \sum_{c\in\mathcal{M}_3'}
   \left[\star j_5-\frac{\mathrm{e}^2}{4\pi^2}(\mathrm{CS})\right]
   \right\}
   \right\rangle_{\mathrm{s}}^B,\qquad
   \delta z=0.
\label{eq:(2.28)}
\end{align}

Finally, we set
\begin{equation}
   \alpha=\frac{2\pi p}{N},
\label{eq:(2.29)}
\end{equation}
where $p$ and~$N$ are coprime integers and define the symmetry operator of the
axial $U(1)$ non-invertible symmetry by~\cite{Honda:2024yte,Honda:2024sdz}
\begin{equation}
   U_{2\pi p/N}(\mathcal{M}_3)
   :=\exp\left\{
   \frac{i\pi p}{N}\sum_{c\in\mathcal{M}_3}
   \left[\star j_5
   -\frac{\mathrm{e}^2}{4\pi^2}(\mathrm{CS})
   \right]\right\}
   \mathcal{Z}_{\mathcal{M}_3}[z],
\label{eq:(2.30)}
\end{equation}
where $\mathcal{Z}_{\mathcal{M}_3}[z]$ is the partition function of a lattice 3D
TQFT. In~Refs.~\cite{Honda:2024yte,Honda:2024sdz}, an explicit form
of~$\mathcal{Z}_{\mathcal{M}_3}[z]$ on the lattice is given by a level~$N$ BF
theory. The basic property of~$\mathcal{Z}_{\mathcal{M}_3}[z]$, for the
configuration~in~Eq.~\eqref{eq:(2.26)} is,
for~$\delta z=0$~\cite{Honda:2024yte},
\begin{equation}
   \mathcal{Z}_{\mathcal{M}_3}[z]
   =\exp\left[
   \frac{i\pi p}{N}\mathrm{e}^2
   \sum_{h\in\mathcal{V}_4}P_2(z)\right]
   \mathcal{Z}_{\mathcal{M}_3'}[z].
\label{eq:(2.31)}
\end{equation}
Combining Eqs.~\eqref{eq:(2.28)} and~\eqref{eq:(2.31)}, we thus have for the
symmetry operator~\eqref{eq:(2.30)},
\begin{equation}
   \left\langle
   U_{2\pi p/N}(\mathcal{M}_3)
   \right\rangle_{\mathrm{s}}^B
   =\left\langle
   U_{2\pi p/N}(\mathcal{M}_3')
   \right\rangle_{\mathrm{s}}^B.
\label{eq:(2.32)}
\end{equation}

We may integrate both sides of~Eq.~\eqref{eq:(2.32)} by $a$ and~$z$ to
yield the full expectation value. Then, we may shift the integration
variable~$z$ by~$z\to z+B$ to make the presence of the 't~Hooft line explicit.
Recalling the relation~\eqref{eq:(2.18)}, we have
\begin{equation}
   \left\langle
   U_{2\pi p/N}(\mathcal{M}_3)T_{\mathrm{q}}(\gamma)
   \right\rangle
   =\left\langle
   U_{2\pi p/N}(\mathcal{M}_3')T_{\mathrm{q}}(\gamma)
   \right\rangle,
\label{eq:(2.33)}
\end{equation}
where the full expectation value is defined by:\footnote{%
We understand
\begin{equation}
   \int\mathrm{D}[a]\mathrm{D}[z]
   :=\prod_{l\in\Gamma}\left[\int_{-\pi}^\pi\mathrm{d}a(l)\right]
   \prod_{p\in\Gamma}\left[\sum_{z(p)=-\infty}^\infty\right].
\label{eq:(2.34)}
\end{equation}
}
\begin{equation}
   \left\langle\dotsb\right\rangle
   :=\int\mathrm{D}[a]\mathrm{D}[z]\mathrm{D}[\Tilde{a}]
   \mathrm{D}[\phi]\mathrm{D}[\ell]
   \mathrm{D}[\chi]\,e^{-S}\dotsb.
\label{eq:(2.35)}
\end{equation}
We note that the shift~$z\to z+B$ does not affect the symmetry operators
in~Eq.~\eqref{eq:(2.32)}, because the defect~$B$ does not have the support
on the 3D boundaries~$\mathcal{M}_3$ and~$\mathcal{M}_3'$ on which the symmetry
operators are defined. Equation~\eqref{eq:(2.33)} and~Fig.~\ref{fig:1} show
that the sweep of the non-invertible symmetry operator over the 't~Hooft line
operator leaves no effect.

Let us summarize the essential elements in our argument. First, we assumed that
the non-genuine but gauge invariant 't~Hooft line operator with an integer
magnetic charge can be realized by the defect~$B$~\eqref{eq:(2.16)}
introduced in the gauge field kinetic term. Next, we assumed that the 4D
topological part in the axial anomaly~$P_2(z)$ with the fractional
coefficient can be represented by a 3D TQFT as in~Eq.~\eqref{eq:(2.31)}; this
enables one to define the symmetry operator as~Eq.~\eqref{eq:(2.30)}. Then,
we have the relation~\eqref{eq:(2.33)}. As far as these assumptions are
fulfilled, the above conclusion follows almost independently of the details of
the lattice action (i.e., possible modifications of~$j_5$, $(\mathrm{CS})$,
$P_2(z)$, and the gauge transformation laws).

\subsection{Axion string}
\label{sec:2.5}
For the axion string, the argument is quite parallel. We introduce a 2D
defect~$\sigma$ by the substitution, $\partial\phi\to\partial\phi-2\pi C$, in
the scalar kinetic term as
\begin{align}
   S^C&:=\sum_{h\in\Gamma}\biggl\{
   \frac{1}{g_0^2}f\cup\star f+i\Tilde{a}\cup\delta z
   +\frac{\mu^2}{2}\left(\partial\phi-2\pi C\right)
   \cup\star\left(\partial\phi-2\pi C\right)
   +i\delta\ell\cup\chi
\notag\\
   &\qquad\qquad{}
   -i\frac{\mathrm{e}^2}{8\pi^2}\left[
   \partial\phi\cup(\mathrm{CS})-4\pi^2\phi\cup P_2(z)\right]
   \biggr\},
\label{eq:(2.36)}
\end{align}
where, assuming $\sigma=\partial\mathcal{V}_3$, we set
\begin{equation}
   C:=-\mathrm{q}\delta_1[\mathcal{V}_3],\qquad
   \delta C=-\mathrm{q}\delta_2[\sigma],
\label{eq:(2.37)}
\end{equation}
where $\delta_1[\mathcal{V}_3]$ is the 1-cochain delta function whose support
is~$\mathcal{V}_3$. Under the change of variable, $\ell\to\ell+C$, we have
\begin{equation}
   e^{-S^C}\to e^{-S}S_{\mathrm{q}}(\sigma),
\label{eq:(2.38)}
\end{equation}
where the axion string operator~$S_{\mathrm{q}}(\sigma)$ is given by
\begin{equation}
   S_{\mathrm{q}}(\sigma)
   =\exp\left\{
   -i\sum_{h\in\Gamma}
   \delta C\cup\left[\chi+\frac{\mathrm{e}^2}{4\pi}a\cup a\right]
   +i\sum_{h\in\Gamma}
   \frac{\mathrm{e}^2}{4\pi}
   C\cup\left(f\cup a+2\pi a\cup z\right)
   \right\},
\label{eq:(2.39)}
\end{equation}
where we have used $\delta z=0$, which is imposed under the integration
over~$\Tilde{a}$. This expression with~Eq.~\eqref{eq:(2.37)} precisely
coincides with that of the axion string operator in~Ref.~\cite{Honda:2024sdz}.

We then introduce the expectation value,
\begin{equation}
   \left\langle\dotsb\right\rangle_{\mathrm{s}}^C
   :=\int\mathrm{D}[\Tilde{a}]\mathrm{D}[\phi]\mathrm{D}[\ell]
   \mathrm{D}[\chi]\,e^{-S^C}\dotsb.
\label{eq:(2.40)}
\end{equation}
Then, under the shift~\eqref{eq:(2.22)},
\begin{equation}
   \left\langle\dotsb\right\rangle_{\mathrm{s}}^C
   =\left\langle
   \exp\left(
   \frac{i}{2}\alpha
   \left\{\sum_{c\in\partial\mathcal{V}_4}
   \left[\star j_5-\frac{\mathrm{e}^2}{4\pi^2}(\mathrm{CS})\right]
   -\mathrm{e}^2\sum_{h\in\mathcal{V}_4}P_2(z)\right\}\right)
   \dotsb\right\rangle_{\mathrm{s}}^C,
\label{eq:(2.41)}
\end{equation}
where we have noted that although the axial $U(1)$ 1-cochain current is
modified as~$j_5\to-2i\mu^2\left(\partial\phi-2\pi C\phi\right)$ in the
presence of the background~$C$, since we assume that the support of~$C$,
$\mathcal{V}_3$, does not touch upon the boundary of the 3D
region~$\partial\mathcal{V}_4$, we can neglect this modification
in~Eq.~\eqref{eq:(2.41)}. Then, repeating the same argument as that for the
't~Hooft line operator, we have
\begin{equation}
   \left\langle
   U_{2\pi p/N}(\mathcal{M}_3)
   \right\rangle_{\mathrm{s}}^C
   =\left\langle
   U_{2\pi p/N}(\mathcal{M}_3')
   \right\rangle_{\mathrm{s}}^C.
\label{eq:(2.42)}
\end{equation}
and, in the full expectation value~\eqref{eq:(2.35)}, after the shift of the
integration variable~$\ell\to\ell+C$, we have
\begin{equation}
   \left\langle
   U_{2\pi p/N}(\mathcal{M}_3)S_{\mathrm{q}}(\sigma)
   \right\rangle
   =\left\langle
   U_{2\pi p/N}(\mathcal{M}_3')S_{\mathrm{q}}(\sigma)
   \right\rangle.
\label{eq:(2.43)}
\end{equation}
Again, the sweep of the symmetry operator over the axion string does not leave
any effect.

Summarizing this paper, we have shown the same conclusions
as~Ref.~\cite{Honda:2024sdz} on the action of the non-invertible symmetry
operator in the axion QED on magnetic objects (the 't~Hooft line and the axion
string) in a much simpler argument.

\section*{Note added}
In this paper and in~Ref.~\cite{Honda:2024sdz}, we have considered the
situation in which one employs only degrees of freedom in the axion QED. The
gauge invariance of line and surface operators then forces us to consider
non-genuine operators. If we regard the axion QED as a low-energy effective
theory of the QED with the Peccei--Quinn scalar, however, there arises a
natural possibility that there exist massless degrees of freedom residing only
on line and surface operators and they make operators genuine. Then, the action
of the non-invertible symmetry operator can be different. We are grateful to
Yuya Tanizaki for helpful discussions on these points. In Appendix~\ref{sec:A},
we give an example of a gauge invariant genuine 't~Hooft line operator on the
lattice and show that with that definition the sweep of the symmetry operator
produces a surface integral of the $U(1)$ field strength as discussed
in~Refs.~\cite{Choi:2022jqy,Cordova:2022ieu,Choi:2022fgx}.

\section*{Acknowledgment}
This work was partially supported by Japan Society for the Promotion of Science
(JSPS) Grant-in-Aid for Scientific Research Grant Number JP23K03418~(H.S.).

\appendix
\section{A possible genuine 't~Hooft line operator in the axion QED}
\label{sec:A}
With only the field contents of the axion QED, it is probably impossible to
construct a gauge invariant 't~Hooft line operator that is really genuine as
we have observed in~Ref.~\cite{Honda:2024sdz} and in this paper. As discussed
in~Ref.~\cite{Choi:2022fgx}, however, if one introduces new degrees of freedom
residing only on the line~$\gamma$, it is possible to construct a gauge
invariant genuine 't~Hooft line operator without affecting the bulk dynamics of
the axion QED. We may adopt, e.g.
\begin{align}
   &\mathcal{T}_{\mathrm{q}}(\gamma)
   :=\exp\left[i\mathrm{q}\sum_{l\in\gamma}
   \left(\Tilde{a}-\frac{\mathrm{e}^2}{2\pi}\phi\cup a\right)
   \right]
\notag\\
   &{}
   \times
   \int\mathrm{D}[\sigma]\mathrm{D}[n]\mathrm{D}[\varphi]\,
   \exp\left[\sum_{l\in\gamma}\left(
   -\frac{\ell_\sigma}{2}D_a\sigma\cup\star D_a\sigma
   -\frac{\ell_\varphi}{2}\partial\varphi\cup\star\partial\varphi
   +\frac{i}{2\pi}\varphi\cup D_a\sigma
   -i\ell\cup\sigma
   \right)\right],
\label{eq:(A1)}
\end{align}
where we have introduced an $\mathbb{R}$ 0-cochain~$\sigma$, a $\mathbb{Z}$
1-cochain~$n$, and another $\mathbb{R}$ 0-cochain~$\varphi$, all are residing
only on the line~$\gamma$ ($\ell_\sigma$ and~$\ell_\varphi$ are
constants).\footnote{The physical origin of such massless degrees of freedom
along the line~$\gamma$ could be attributed to fermion zero modes in QED
localized on the monopole worldline~\cite{Yamagishi:1982wp}. We would like to
thank Yuya Tanizaki for enlightening us as to this point.} In this expression,
the covariant derivatives are defined by
\begin{equation}
   D_a\sigma:=\delta\sigma+2\pi n+\mathrm{q}\mathrm{e}^2a,\qquad
   \partial\varphi:=\delta\varphi+2\pi\ell.
\label{eq:(A2)}
\end{equation}
In conjunction with the gauge transformations, Eqs.~\eqref{eq:(2.7)},
\eqref{eq:(2.8)}, \eqref{eq:(2.9)}, \eqref{eq:(2.12)}, and~\eqref{eq:(2.13)},
we define gauge transformations of $\sigma$, $n$, and~$\varphi$ by
\begin{equation}
   \sigma\to\sigma-\mathrm{q}\mathrm{e}^2\lambda,\qquad
   n\to n-\mathrm{q}\mathrm{e}^2m,\qquad
   \varphi\to\varphi+2\pi k.
\label{eq:(A3)}
\end{equation}
Then, noting that the covariant derivatives~\eqref{eq:(A2)} are gauge
invariant, it is straightforward to see that the line
operator~$\mathcal{T}_{\mathrm{q}}(\gamma)$ in~Eq.~\eqref{eq:(A1)} is invariant
under all gauge transformations.\footnote{%
The dual gauge potential in~Ref.~\cite{Choi:2022fgx} corresponds to our
$\Tilde{a}-\mathrm{e}^2/(2\pi)\phi\cup a$. We emphasize however that the line
operator~\eqref{eq:(A1)} is \emph{not\/} a lattice transcription of the 1D
theories residing on the line considered in~Ref.~\cite{Choi:2022fgx}. A natural
lattice transcription of~Eq.~(5.2) of~Ref.~\cite{Choi:2022fgx} would be
\begin{equation}
   \Check{\mathcal{T}}_{\mathrm{q}}(\gamma)
   :=\exp\left[i\mathrm{q}\sum_{l\in\gamma}
   \left(\Tilde{a}-\frac{\mathrm{e}^2}{2\pi}\phi\cup a\right)
   \right]
   \int\mathrm{D}[\sigma]\mathrm{D}[n]\,
   \exp\left[\sum_{l\in\gamma}\left(
   -\frac{\ell_\sigma}{2}D_a\sigma\cup\star D_a\sigma
   +\frac{i}{2\pi}\phi\cup D_a\sigma
   -i\ell\cup\sigma
   \right)\right].
\label{eq:(A4)}
\end{equation}
This line operator, however, does not exhibit a simple transformation law under
the sweep of the non-invertible symmetry operator~\eqref{eq:(2.30)}. Repeating
the argument in this Appendix for this operator, we have
\begin{align}
   &\left\langle
   U_{2\pi p/N}(\mathcal{M}_3)\Check{\mathcal{T}}_{\mathrm{q}}(\gamma)
   \right\rangle
\notag\\
   &=\Biggl\langle
   \exp
   \left(
   -\frac{ip}{N}\mathrm{q}\mathrm{e}^2\sum_{p\in\mathcal{R}}f
   \right)
   U_{2\pi p/N}(\mathcal{M}_3')
\notag\\
   &\qquad{}
   \times\exp\left[i\mathrm{q}\sum_{l\in\gamma}
   \left(\Tilde{a}-\frac{\mathrm{e}^2}{2\pi}\phi\cup a\right)
   \right]
\notag\\
   &\qquad{}
   \times
   \int\mathrm{D}[\sigma]\mathrm{D}[n]\,
   \exp\left[\sum_{l\in\gamma}\left(
   -\frac{\ell_\sigma}{2}D_a\sigma\cup\star D_a\sigma
   +\frac{i}{2\pi}\phi\cup D_a\sigma
   -i\ell\cup\sigma
   \right)\right]e^{i\frac{p}{N}\sum_{l\in\gamma}D_a\sigma}
   \Biggr\rangle.
\label{eq:(A5)}
\end{align}
Since the last factor~$e^{i\frac{p}{N}\sum_{l\in\gamma}D_a\sigma}$ depends on the
winding number of the compact scalar~$\sigma$ on~$\gamma$, $\sum_{l\in\gamma}n$,
the factor~$e^{i(p/N)\sum_{l\in\gamma}D_a\sigma}$ cannot simply be pulled out from the
functional integral over~$\sigma$ and~$n$; thus the line operator does not keep
its original structure under the sweep of the symmetry operator. Note that the
quantity in the right-hand side is still gauge invariant.}

To derive the axial Ward--Takahashi identity for the line operator
$\mathcal{T}_{\mathrm{q}}(\gamma)$~\eqref{eq:(A1)}, we take
\begin{equation}
   \exp\left\{
   i\frac{\pi p}{N}
   \sum_{c\in\mathcal{M}_3}
   \left[\star j_5-\frac{\mathrm{e}^2}{4\pi^2}(\mathrm{CS})\right]
   \right\}
   \mathcal{T}_{\mathrm{q}}(\gamma),
\label{eq:(A6)}
\end{equation}
for $\dotsb$ in~Eq.~\eqref{eq:(2.21)}; in this Appendix, we set~$B=0$. Under the
chiral transformation~\eqref{eq:(2.22)}, the lattice action changes
by~Eq.~\eqref{eq:(2.23)} (where~$\alpha=2\pi p/N$). In the last term
of~Eq.~\eqref{eq:(2.23)}, we use
\begin{align}
   P_2(z)&=z_0\cup z_0+2z_0\cup z'+\delta(z_0\cup_1z')
   +z'\cup z'+z'\cup_1\delta z'
\notag\\
   &=z_0\cup z_0+2\mathrm{q}z_0\cup\delta_2[\mathcal{R}]
   +\delta(z_0\cup_1z'),
\label{eq:(A7)}
\end{align}
for $z$ in the presence of the 't~Hooft line~$\gamma=\partial\mathcal{R}$,
\begin{equation}
   z=z_0+z',\qquad\delta z_0=0,\qquad z':=\mathrm{q}\delta_2[\mathcal{R}].
\label{eq:(A8)}
\end{equation}
On the other hand, under~Eq.~\eqref{eq:(2.22)}, Eq.~\eqref{eq:(A6)} changes
into
\begin{align}
   &\exp\left\{
   i\frac{p}{N}
   \sum_{c\in\mathcal{M}_3}
   \left[\star j_5-\frac{\mathrm{e}^2}{4\pi^2}(\mathrm{CS})\right]
   \right\}
   \exp\left(-i\frac{p}{N}\mathrm{q}\mathrm{e}^2\sum_{l\in\gamma}a\right)
   \mathcal{T}_{\mathrm{q}}(\gamma).
\label{eq:(A9)}
\end{align}
Finally, noting that Eq.~\eqref{eq:(2.31)} implies
\begin{equation}
   \mathcal{Z}_{\mathcal{M}_3}[z]
   =\exp\left(
   \frac{i\pi p}{N}\mathrm{e}^2
   \sum_{h\in\mathcal{V}_4}z_0\cup z_0\right)
   \mathcal{Z}_{\mathcal{M}_3'}[z],
\label{eq:(A10)}
\end{equation}
and recalling the definition of the symmetry
operator~$U_{2\pi p/N}(\mathcal{M}_3)$ in~Eq.~\eqref{eq:(2.30)}, we have the
identity
\begin{align}
   \left\langle
   U_{2\pi p/N}(\mathcal{M}_3)\mathcal{T}_{\mathrm{q}}(\gamma)
   \right\rangle
   &=\left\langle
   \exp
   \left(
   -i\frac{2\pi p}{N}\mathrm{q}\mathrm{e}^2\sum_{p\in\mathcal{R}}z
   -i\frac{p}{N}\mathrm{q}\mathrm{e}^2\sum_{l\in\gamma}a\right)
   U_{2\pi p/N}(\mathcal{M}_3')\mathcal{T}_{\mathrm{q}}(\gamma)
   \right\rangle
\notag\\
   &=\left\langle
   \exp
   \left(
   -i\frac{p}{N}\mathrm{q}\mathrm{e}^2\sum_{p\in\mathcal{R}}f\right)
   U_{2\pi p/N}(\mathcal{M}_3')\mathcal{T}_{\mathrm{q}}(\gamma)
   \right\rangle.
\label{eq:(A11)}
\end{align}
This shows that the sweep of the non-invertible symmetry operator over the
genuine 't~Hooft line operator defined by~Eq.~\eqref{eq:(A1)} produces the
surface integral of the field strength~$f$, as discussed
in~Refs.~\cite{Choi:2022jqy,Cordova:2022ieu,Choi:2022fgx}.

\let\doi\relax









\end{document}